\documentclass{llncs}

\usepackage{amsmath}
\usepackage{comment}
\usepackage{amsmath}
\usepackage{amssymb}
\usepackage{amsfonts}
\usepackage{graphicx}
\usepackage{subfig}
\usepackage{color}
\usepackage{multicol}
\usepackage{multirow}
\newcommand{\bumpup}{\vspace*{-2.0ex}}
\usepackage[ruled,noline]{algorithm2e}
\usepackage{comment}
\usepackage{float}

\usepackage{wrapfig}

\begin{document}



\title{Self-supervised 3D anatomy segmentation using self-distilled masked image transformer (SMIT)} 

\author{Jue Jiang\inst{1}, Neelam Tyagi\inst{1}, Kathryn Tringale \inst{2}, Christopher Crane \inst{2},
  \and Harini Veeraraghavan \inst{1}}


\institute{\inst{1} Department of Medical Physics, Memorial Sloan Kettering Cancer Center \and %
                      \inst{2} Department of Radiation Oncology, Memorial Sloan Kettering Cancer Center
           \email{veerarah@mskcc.org}       
              }

\maketitle              
\begin{abstract}
Vision transformers, with their ability to more efficiently model long-range context, have demonstrated impressive accuracy gains in several computer vision and medical image analysis tasks including segmentation. However, such methods need large labeled datasets for training, which is hard to obtain for medical image analysis. Self-supervised learning (SSL) has demonstrated success in medical image segmentation using convolutional networks. In this work, we developed a \underline{s}elf-distillation learning with \underline{m}asked \underline{i}mage modeling method to perform SSL for vision \underline{t}ransformers (SMIT) applied to 3D multi-organ segmentation from CT and MRI. Our contribution is a dense pixel-wise regression within masked patches called masked image prediction, which we combined with masked patch token distillation as pretext task to pre-train vision transformers. We show our approach is more accurate and requires fewer fine tuning datasets than other pretext tasks. Unlike prior medical image methods, which typically used image sets arising from disease sites and imaging modalities corresponding to the target tasks, we used 3,643 CT scans (602,708 images) arising from head and neck, lung, and kidney cancers as well as COVID-19 for pre-training and applied it to abdominal organs segmentation from MRI pancreatic cancer patients as well as publicly available 13 different abdominal organs segmentation from CT. Our method showed clear accuracy improvement (average DSC of 0.875 from MRI and 0.878 from CT) with reduced requirement for fine-tuning datasets over commonly used pretext tasks. Extensive comparisons against multiple current SSL methods were done. Code will be made available upon acceptance for publication. \footnote{\textcolor{blue}{This paper has been early accepted by MICCAI 2022.}} 
 

\end{abstract}
\bumpup
\bumpup
\section{Introduction}
Vision transformers (ViT)\cite{dosovitskiy2021} efficiently model long range contextual information using multi-head self attention mechanism, which makes them robust to occlusions, image noise, as well as domain and image contrast differences. ViTs have shown impressive accuracy gains over convolutional neural networks (CNN) in medical image segmentation\cite{Xie_CoTr_MICCAI21,hatamizadeh2021unetr}. However, ViT training requires a large number of expert labeled training datasets that are not commonly available in medical image applications. Self-supervised learning (SSL) overcomes the requirement for large labeled training datasets by using large unlabeled datasets through pre-defined annotation free pretext tasks. The pretext tasks are based on modeling visual information contained in images and provide a surrogate supervision signal for feature learning\cite{noroozi2016unsupervised,komodakis2018unsupervised,he2020momentum}. Once pre-trained, the model can be re-purposed for a variety of tasks and require relatively few labeled sets for fine-tuning. 
\\ 
The choice of pretext tasks is crucial to successfully mine useful image information using SSL. Pretext tasks in medical image applications typically focus on learning denoising autoencoders constructed with CNNs to recover images in the input space using corrupted versions of the original images\cite{zhou2021models,haghighi2020learning}. Various data augmentation strategies have been used to corrupt images, which include jigsaw puzzles\cite{taleb20203d,zhu2020rubik}, transformation of image contrast and local texture\cite{zhou2021models}, image rotations\cite{komodakis2018unsupervised}, and masking of whole image slices\cite{jun2021medical}. Learning strategies include pseudo labels\cite{haghighi2020learning,chen2019self,sun2021unsupervised} and contrastive learning\cite{taleb20203d,chaitanya2020contrastive,feng2020parts2whole,zhou2021preservational}. However, CNNs are inherently limited in their capacity to model long-range context than transformers, which may reduce their robustness to imaging variations and contrast differences. Hence, we combined ViT with SSL using masked image modeling (MIM) and self-distillation of concurrently trained teacher and student networks.
\\
MIM has been successfully applied to transformers to capture local context while preserving global semantics in natural image analysis tasks\cite{li2021mst,xie2021simmim,zhou2022image,bao2021beit,zhou2022image,he2021masked}. Knowledge distillation with concurrently trained teacher has also been used for medical image segmentation by leveraging different imaging modality datasets (CT and MRI)\cite{Li_Yu_Wang_Heng_2020,jiang2021unpaired}. Self-distillation on the other hand, uses different augmented views of the same image\cite{caron2021emerging} and has been used with contrastive learning with convolutional encoders for medical image classification\cite{sun2021unsupervised}.
\\
Self-distillation learning with MIM and using a pair of online teacher and a student transformer encoders have been used for natural image classification and segmentation\cite{zhou2022image,caron2021emerging}. However, the pretext tasks focused only on extracting global image embedding as class tokens [CLS]\cite{caron2021emerging}, which was improved with global and local patch token embeddings \cite{zhou2022image}. However, these methods ignored the dense pixel dependencies, which is essential for dense prediction tasks like segmentation. Hence, we introduced a masked image prediction (MIP) pretext task to predict pixel-wise intensities within masked patches combined with the local and global embedding distillation applied to medical image segmentation. 
\underline{Our contributions include\/\rm:} (i) SSL using MIM and self-distillation approach combining masked image prediction, masked patch token distillation, and global image token distillation for CT and MRI organs segmentation using transformers. (ii) a simple linear projection layer for medical image reconstruction to speed up pre-training, which we show is more accurate than multi-layer decoder. (iii) SSL pre-training using large 3,643 3D CTs arising from a variety of disease sites including head and neck, chest, and abdomen with different cancers (lung, naso/oropharynx, kidney) and COVID-19 applied to CT and MRI segmentation. (iv) Evaluation of various pretext tasks using transformer encoders related to fine tuning data size requirements and segmentation accuracy.

	\begin{figure*}[tpb]
		\begin{center}
			\includegraphics[width=0.9\columnwidth,scale=1]{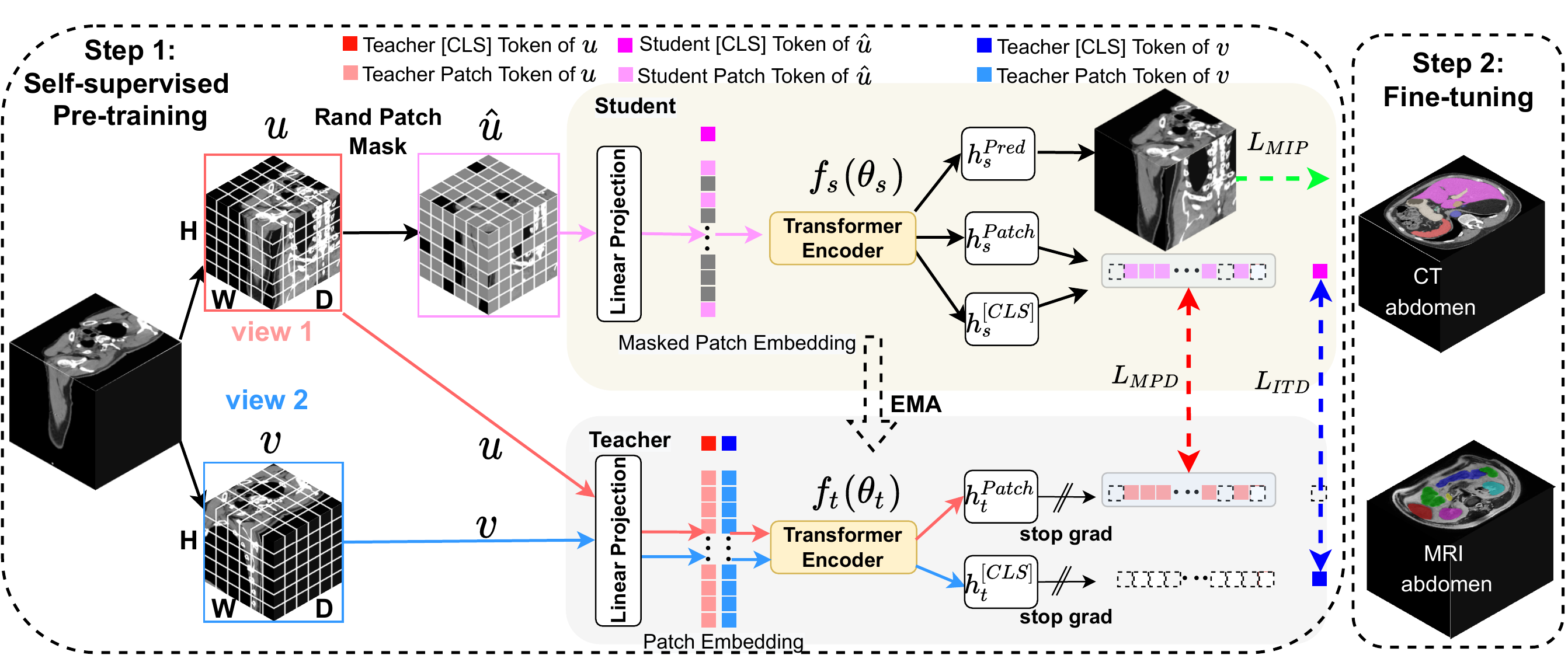} 
			\vspace{-0.05cm}\setlength{\belowcaptionskip}{-0.6cm}\setlength{\abovecaptionskip}{0.08cm}\caption{\small SMIT: \underline{S}elf-distillation with \underline{m}asked \underline{i}mage modeling for \underline{t}ransformers using SSL. }\label{fig:method}
		\end{center}
	\end{figure*}

\section{Method}
\bumpup
\textbf{Goal\/\rm:}  Extract a universal representation of images for dense prediction tasks, given an unlabeled dataset of $Q$ images. 
\\
\textbf{Approach\/\rm:} A visual tokenizer $f_{s}(\theta_{s})$ implemented as a transformer encoder is learned via self-distillation using MIM pretext tasks in order to convert an image $x$ into image tokens $\{x_{i}\}_{i=1}^{N}$, $N$ being the sequence length. Self distillation is performed by concurrently training an online teacher tokenizer model $f_{t}(\theta_{t})$ with the same network structure as $f_s(\theta_{s})$ serving as the student model. A global image token distillation (ITD) is performed as a pretext task to match the global tokens extracted by $f_t$ and $f_s$ as done previously\cite{caron2021emerging}. MIM pretext tasks include masked image prediction (MIP) and masked patch token distillation (MPD). 
\\
Suppose $\{u,v\}$ are two augmented views of a 3D image $x$. $N$ image patches are extracted from the images to create a sequence of image tokens\cite{dosovitskiy2021}, say $u = \{u_i\}_{i=1}^{N}$. The image tokens are then corrupted by randomly masking image tokens based on a binary vector $m = \{m_{i}\}_{i=1}^{N} \in \{0,1\}$ with a probability $p$ and then replacing with mask token\cite{bao2021beit} $e_{[MASK]}$ such that as $\tilde{u} = m\odot u$ with $\tilde{u_i} = e_{[MASK]}$ at $m_i = 1$ and $\tilde{u_i} = u_i$ at $m_i = 0$. The second augmented view $v$ is also corrupted but using a different mask vector instance $ m^{\prime}$ as $\tilde{v} = m^{\prime} \odot v$. \\
\textbf{Dense pixel dependency modeling using MIP: \/}\rm MIP involves recovering the original image view $u$ from corrupted $\tilde{u}$, as $\hat{u}=h_{s}^{Pred}(f_{s}(\tilde{u}, \theta_{s}))$, where $h_{s}^{Pred}$ decodes the visual tokens produced by a visual tokenizer $f_{s}(\theta_{s})$ into images (see Fig.\ref{fig:method}). MIP involves dense pixel regression of image intensities within masked patches using the context of unmasked patches. The MIP loss is computed as (dotted green arrow in Fig.\ref{fig:method}):
\begin{equation}
\setlength{\abovedisplayskip}{1pt}
\setlength{\belowdisplayskip}{1pt} 
L_{MIP} = \sum_{i}^{N} E\| m_{i} \cdot (h_{s}^{Pred}(f_{s}(\tilde{u_{i}}, \theta_{s}))) - u_{i}) \|_{1} 
\end{equation}  
$h_s^{Pred}$ is a linear projection with one layer for dense pixel regression. A symmetrized loss using $v$ and $\tilde{v}$ is  combined to compute the total loss for $L_{MIP}$.
\\
\textbf{Masked patch token self-distillation (MPD): \/}\rm MPD is accomplished by optimizing a teacher $f_{t}(\theta_{t})$ and a student visual tokenizer $f_{s}(\theta_{s})$ such that the student network predicts the tokens of the teacher network. The student network $f_s$ tokenizes the corrupted version of an image $\tilde{u}$ to generate visual tokens $\phi^{\prime} = \{\phi^{\prime}_{i}\}_{i=1}^{N}$. The teacher network $f_t$ tokenizes the uncorrupted version of the same image $u$ to generate visual tokens $\phi = \{\phi_{i}\}_{i=1}^{N}$. Similar to MIP, MPD is only concerned with ensuring prediction of the masked patch tokens. Therefore, the loss is computed from masked portions (i.e. $m_i$=1) using cross-entropy of the predicted patch tokens (dotted red arrow in Fig.\ref{fig:method}): 
\begin{equation}
\setlength{\abovedisplayskip}{1pt}
\setlength{\belowdisplayskip}{1pt} 
\begin{split}
L_{MPD} = - \sum_{i=1}^N m_i \cdot  P_{t}^{Patch}(u_{i},\theta_t) log (P_{s}^{Patch}(\tilde{u}_{i},\theta_s)),
\end{split}
\end{equation}
where $P_s^{Patch}$ and $P_t^{Patch}$ are the patch token distributions for student and teacher networks. They are computed by applying \it{softmax \/}\rm to the outputs of $h_s^{Patch}$ and $h_t^{Patch}$. The sharpness of the token distribution is controlled using a temperature term $\tau_{s} >0$ and $\tau_{t} > 0$ for the student and teacher networks, respectively. Mathematically, such a sharpening can expressed as (using notation for the student network parameters) as:
\begin{equation}
\setlength{\abovedisplayskip}{1pt}
\setlength{\belowdisplayskip}{1pt} 
    P_{s}^{Patch}(u,\theta_{s}) = \frac{exp(h_s^{Patch}(f_{s}(u_{j},\theta_{s}))/\tau_{s}}{\sum_{j=1}^{K}exp(h_s^{Patch}(f_{s}(u_{j},\theta_{s}))/\tau_{s}}.
\end{equation}
A symmetrized cross entropy loss corresponding to the other view $v$ and $\tilde{v}$ is also computed and averaged to compute the total loss for MPD.
\\
\textbf{Global image token self-distillation (ITD): \/}\rm ITD is done by matching the global image embedding represented as class tokens [CLS] distribution $P_{s}^{[CLS]}$ extracted from the corrupted view $\tilde{u}$ by student using $h_s^{[CLS]}(f_{s}(\theta_{s},\tilde{u}))$ with the token distribution $P_{t}^{[CLS]}$ extracted from the uncorrupted and different view $v$ by the teacher using $h_{t}^{[CLS]}(f_{t}(\theta_{t}, v))$ (shown by dotted blue arrow in Fig.\ref{fig:method}) as:  
\begin{equation}
\setlength{\abovedisplayskip}{1pt}
\setlength{\belowdisplayskip}{1pt} 
\begin{split}
L_{ITD} = - \sum_{i=1}^N m_i \cdot P_{t}^{[CLS]}(v_{i},\theta_t) log (P_{s}^{[CLS]}(\tilde{u}_{i},\theta_s))
\end{split}
\end{equation}
Sharpening transforms are applied to $P_t^{[CLS]}$ and $P_s^{[CLS]}$ similar to Equation 4. A symmetrized cross entropy loss corresponding to the corrupted view $\tilde{v}$ and another $u$ is also computed and averaged to compute the total loss for $L_{ITD}$.\\
\textbf{Online teacher network update: \/}\rm Teacher network parameters were updated using exponential moving average (EMA) with momentum update, and shown to be feasible for SSL\cite{caron2021emerging,zhou2022image} as: $\theta_{t}=\lambda_m\theta_{t} + (1-\lambda_m)\theta_{s}$,
where $\lambda_{m}$ is momentum, which was updated using a cosine schedule from 0.996 to 1 during training. The total loss was, $L_{total}$ = $L_{MIP}$ + $\lambda_{MPD}$ $L_{MPD}$ + $\lambda_{ITD}$ $L_{ITD}$.
\\
\textbf{Implementation details: \/} \rm 
All the networks were implemented using the Pytorch library and trained on 4 Nvidia GTX V100. SSL optimization was done using ADAMw with a cosine learning rate scheduler trained for 400 epochs with an initial learning rate of 0.0002 and warmup for 30 epochs. $\lambda_{MPD}$=0.1, $\lambda_{ITD}$ =0.1 were set experimentally. A default mask ratio of 0.7 was used. Centering and sharpening operations reduced chances of degenerate solutions\cite{caron2021emerging}. $\tau_{s}$ was set to 0.1 and $\tau_t$ was linearly warmed up from 0.04 to 0.07 in the first 30 epochs. SWIN-small backbone\cite{liu2021swin} with 768 embedding, window size of 4 $\times$ 4 $\times$ 4, patch size of 2 was used. The 1-layer decoder was implemented with a linear projection layer with the same number of output channels as input image size. The network had 28.19M parameters. Following pre-training, only the student network was retained for fine-tuning and testing.
\bumpup
\section{Experiments and Results}
\bumpup

\subsubsection{Training dataset:} SSL pre-training was performed using 3,643 CT patient scans containing 602,708 images. Images were sourced from patients with head and neck (N$=$837) and lung cancers (N$=$1455) from internal and external\cite{aerts2015data}, as well as those with kidney cancers\cite{akin2016radiology} (N$=$710), and COVID-19\cite{harmon2020artificial} (N$=$650). GPU limitation was addressed for training, fine-tuning, and testing by image resampling (1.5$\times$1.5$\times$2mm voxel size) and cropping (128$\times$128$\times$128) to enclose the body region. Augmented views for SSL training was produced through randomly cropped 96$\times$96$\times$96 volumes, which resulted in 6$\times$6$\times$6 image patch tokens. A sliding window strategy with half window overlap was used for testing\cite{Xie_CoTr_MICCAI21,hatamizadeh2021unetr}.
\begin{table}[tp] 
	\centering{\caption{\small{Accuracy on BTCV standard challenge test set. SP: spleen, RK/LK: right \& left kidney, GB: gall bladder, ESO: esophagus, LV: liver, STO: stomach, AOR: aorta, IVC: inferior vena cava, SPV: portal \& splenic vein, Pan: Pancreas, AG: Adrenals.}} 
	\label{tab:abdomen_result_test} 
	\centering
	
    \scriptsize
	\begin{tabular}{|c|c|c|c|c|c|c|c|c|c|c|c|c|c|}
	\hline 
	{Method} & {SP}& {RK}&{LK} &{GB} &{ESO}& {LV} &{STO}&{AOR}&{IVC}&{SPV}&{Pan}&{AG}&{AVG}\\
	\hline
	{ASPP\cite{chen2018encoder} }&0.935& 0.892& 0.914& 0.689 & 0.760 & 0.953 & 0.812& 0.918 & 0.807& 0.695 & 0.720 & 0.629 &0.811\\
\hline
{nnUnet\cite{isensee2021nnu}}&0.942&0.894&0.910&0.704&0.723&0.948&0.824&0.877&0.782&0.720&0.680&0.616&0.802\\
	\hline
{TrsUnet\cite{chen2021transunet} }&0.952&0.927&0.929&0.662&0.757&0.969&0.889&0.920&0.833&0.791&0.775&0.637&0.838
\\
	\hline	
	{CoTr\cite{Xie_CoTr_MICCAI21} }&0.958&0.921&0.936&0.700&0.764&0.963&0.854&0.920&0.838&0.787&0.775&0.694&0.844\\
	\hline
	{UNETR\cite{hatamizadeh2021unetr} }&\textbf{0.968}&0.924&0.941&0.750&0.766&0.971&0.913&0.890& 0.847 &0.788&0.767&0.741&0.856\\
    \hline	
    {SMIT(rand)}&0.959&{0.921}&	{0.947} &	{0.746}&	{0.802}&	{0.972}&	{0.916} &	{0.917}&	{0.848}&	{0.797} &	{0.817}&	{0.711} &{0.850}\\ 
    {SMIT(SSL)}&0.967&\textbf{0.945}&	\textbf{0.948} &	\textbf{0.826}&	\textbf{0.822}&	\textbf{0.976}&	\textbf{0.934} &	\textbf{0.921}&	\textbf{0.864}&	\textbf{0.827} &	\textbf{0.851}&	\textbf{0.754} &\textbf{0.878}\\    
		\hline	
    \end{tabular}} 
\end{table}	
\\\underline{CT abdomen organ segmentation (Dataset I): \/}\rm The pre-trained networks were fine-tuned to generate volumetric segmentation of 13 different abdominal organs from contrast-enhanced CT (CECT) scans using publicly available beyond the cranial vault (BTCV)\cite{landman2015miccai} dataset. Randomly selected 21 images are used for training and the remaining used for validation. Furthermore, blinded testing of 20 CECTs evaluated on the grand challenge website is also reported.
\\
\underline{MRI upper abdominal organs segmentation (Dataset II): \/}\rm The SSL network was evaluated for segmenting abdominal organs at risk for pancreatic cancer radiation treatment, which included  stomach, small and large bowel, liver, and kidneys. \underline{No MRI or pancreatic cancer scans were used for SSL pre-training\/\rm}. Ninety two 3D T2-weighted MRIs (TR/TE = 1300/87 ms, voxel size of 1$\times$1$\times$2 mm$^{3}$, FOV of 400$\times$450$\times$250 mm$^{3}$) and acquired with pnuematic compression belt to suppress breathing motion were analyzed. Fine tuning used five-fold cross-validation and results from the validation folds not used in training are reported.
	\begin{figure*}[tp]
		\begin{center}
			\includegraphics[width=0.7\columnwidth,scale=0.7]{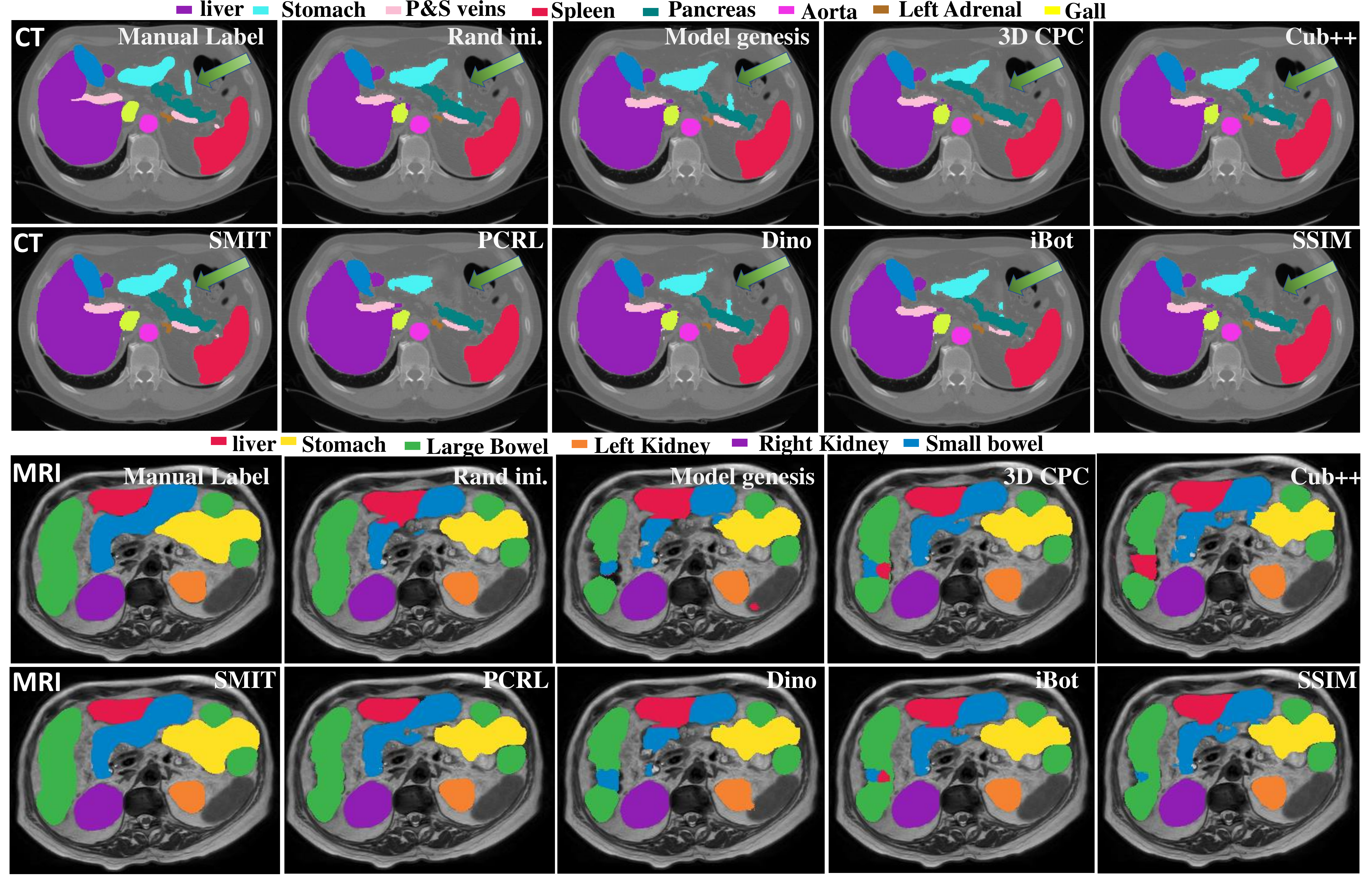}
			\vspace{-0.05cm}\setlength{\belowcaptionskip}{-0.8cm}\setlength{\abovecaptionskip}{0.08cm}\caption{\small Segmentation performance of different methods on MRI abdomen organs.} \label{fig:seg_overlay_MRI}
		\end{center}
	\end{figure*}
\textbf{Experimental comparisons: \/}\rm SMIT was compared against representative SSL medical image analysis methods. Results from representative published methods on the BTCV testing set\cite{isensee2021nnu,Xie_CoTr_MICCAI21,hatamizadeh2021unetr} are also reported. The SSL comparison methods were chosen to evaluate the impact of the pretext task on segmentation accuracy and included (a) local texture and semantics modeling using model genesis\cite{zhou2021models}, (b) jigsaw puzzles\cite{zhu2020rubik}, (c) contrastive learning\cite{zhou2021preservational} with (a),(b), (c) implemented on CNN backbone, (d) self-distillation using whole image reconstruction\cite{caron2021emerging}, (e) masked patch reconstruction\cite{xie2021simmim} without self-distillation, (f) MIM using self-distillation\cite{zhou2022image} with (d),(e), and (f) implemented in a SWIN transformer backbone. Random initialization results are shown for benchmarking purposes using both CNN and SWIN backbones. Identical training and testing sets were used with hyper-parameter adopted from their default implementation. \\
\textbf{CT segmentation accuracy: \/}\rm As shown in Table.\ref{tab:abdomen_result_test}, our method SMIT outperformed representative published methods including transformer-based segmentation\cite{chen2021transunet,hatamizadeh2021unetr,Xie_CoTr_MICCAI21}. SMIT was also more accurate than all evaluated SSL methods (Table.\ref{tab:CT_MRI_result}) for most organs. Prior-guided contrast learning (PRCL)\cite{zhou2021preservational} was more accurate than SMIT for gall bladder (0.797 vs. 0.787). SMIT was more accurate than self-distillation with MIM\cite{zhou2022image} (average DSC of 0.848 vs. 0.833) as well as masked image reconstruction without distillation\cite{xie2021simmim} (0.848 vs. 0.830). Fig.\ref{fig:seg_overlay_MRI} shows a representative case with multiple organs segmentations produced by the various methods. SMIT was the most accurate method including for organs with highly variable appearance and size such as the stomach and esophagus.
\\
\textbf{MRI segmentation accuracy: \/}\rm SMIT was more accurate than all other SSL-based methods for all evaluated organs, including stomach and bowels, which depict highly variable appearance and sizes (Table.\ref{tab:CT_MRI_result}). SMIT was least accurate for small bowel compared to other organs, albeit this accuracy for small bowel was higher than all other methods. Fig.\ref{fig:seg_overlay_MRI} shows a representative case with multiple organs segmentations produced by the various methods.
	\begin{figure*}[tp]
		\begin{center}
			\includegraphics[width=1.0\columnwidth,scale=0.7]{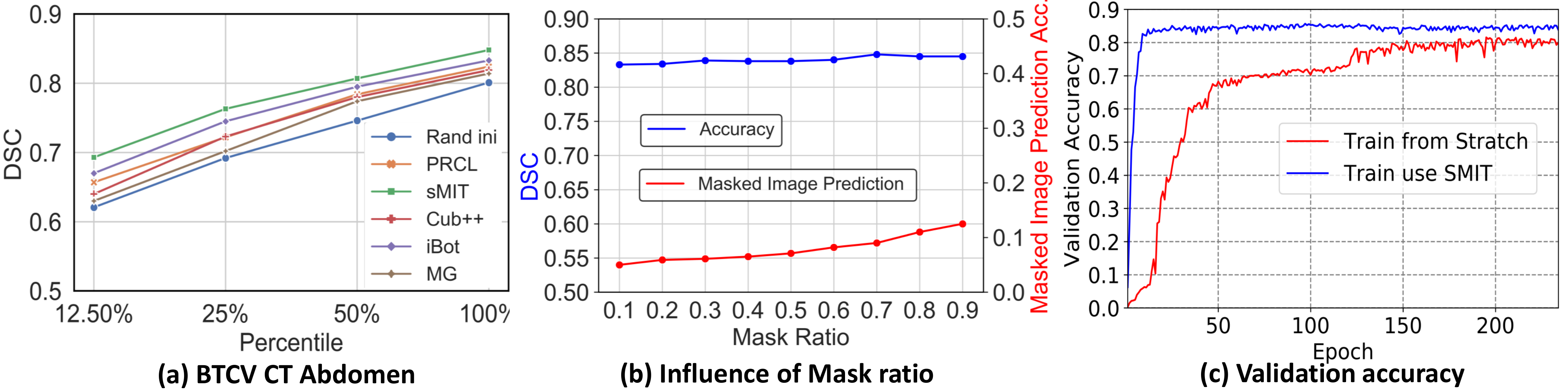}
			\vspace{-0.05cm}\setlength{\belowcaptionskip}{-0.8cm}\setlength{\abovecaptionskip}{0.08cm}\caption{\small (a) Impact of SSL task on fine-tuning sizes, (b) impact of mask ratio on masked image prediction and segmentation accuracy, (c) training convergence.} \label{fig:sample_size}
		\end{center}
	\end{figure*}
\\	
\textbf{Ablation experiments: \/}\rm All ablation and design experiments (1layer decoder vs. multi-layer or ML decoder) were performed using the BTCV dataset and used the SWIN-backbone as used for SMIT. ML decoder was implemented with five transpose convolution layers for up-sampling back to the input image resolution. Fig.\ref{fig:diff_tasks_group} shows the accuracy comparisons of networks pre-trained with different tasks including  full image reconstruction, contrastive losses, pseudo labels\cite{chen2021empirical}, and various combination of the losses ($L_{MIP}, L_{MPD}, L_{ITD}$). As shown, the accuracies for all the methods was similar for large organs depicting good contrast that include liver, spleen, left and right kidney (Fig.\ref{fig:diff_tasks_group}(I)). On the other hand, organs with low soft tissue contrast and high variability (Fig.\ref{fig:diff_tasks_group}(II)) and small organs (Fig.\ref{fig:diff_tasks_group}(III)) show larger differences in accuracies between methods with SMIT achieving more accurate segmentations. Major blood vessels Fig.\ref{fig:diff_tasks_group}(IV) also depict segmentation accuracy differences across methods, albeit less so than for small organs and those with low soft-tissue contrast. Importantly, both full image reconstruction and multi-layer decoder based MIP (ML-MIP) were less accurate than SMIT, which uses masked image prediction with 1-layer linear projection decoder (Fig.\ref{fig:diff_tasks_group} (II,III,IV)). MPD was the least accurate for organs with low soft-tissue contrast and high variability (Fig.\ref{fig:diff_tasks_group}(II)), which was improved slightly by adding global image distillation (ITD). MIP alone (using 1-layer decoder) was similarly accurate as SMIT and more accurate than other pretext task based segmentation including ITD\cite{caron2021emerging}, MPD+ITD\cite{zhou2022image}.
\\
\textbf{Impact of pretext tasks on sample size for fine tuning: \/}\rm SMIT was more accurate than all other SSL methods irrespective of sample size used for fine-tuning (Fig.\ref{fig:sample_size}(a)) and achieved faster convergence (Fig.\ref{fig:sample_size}(c)). It outperformed iBot\cite{zhou2022image}, which uses MPD and ITD, indicating effectiveness of MIP for SSL.
\begin{figure*}[tp]
		\begin{center}
			\includegraphics[width=0.75\columnwidth,scale=0.8]{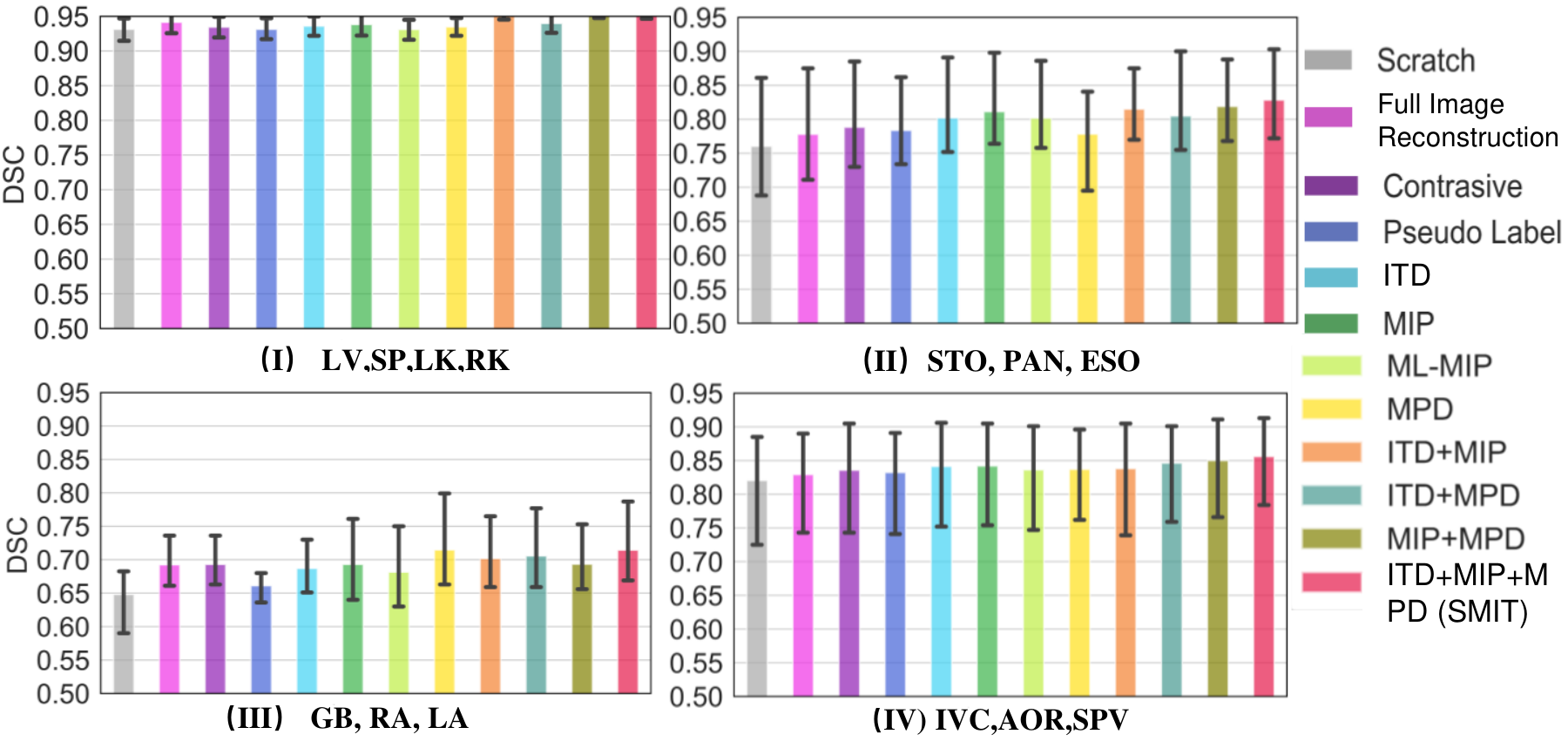}
			\vspace{-0.05cm}\setlength{\belowcaptionskip}{-0.8cm}\setlength{\abovecaptionskip}{0.08cm}\caption{\small Accuracy variations by organ types using different pretext tasks.} \label{fig:diff_tasks_group}
		\end{center}
	\end{figure*}
\textbf{Impact of mask ratio on accuracy: \/}\rm Fig.\ref{fig:sample_size}(b) shows the impact of mask ratio (percentage of masked patches) in the corrupted image for both the accuracy of masked image reconstruction (computed as mean square error [MSE]) as well as segmentation (computed using DSC metric). Increasing the mask ratio initially increased accuracy and then stabilized. Image reconstruction error increased slightly with increasing masking ratio. Fig.\ref{fig:rec_img} shows a representative CT and MRI reconstruction produced using default and multi-layer decoder, wherein our method was more accurate even in highly textured portions of the images containing multiple organs (additional examples are shown in Supplementary Fig 1). Quantitative comparisons showed our method was more accurate (MSE of 0.061 vs. 0.32) for CT (N$=$10 cases) and 92 MRI (MSE of 0.062 vs. 0.34) than multi-layer decoder. 

\begin{table*}[htp]
    \centering{\caption{CT and MRI segmentation accuracy comparisons to SSL methods. Rand-random; LB-Large bowel, SB - Small bowel.} 
	\label{tab:CT_MRI_result} 
	\centering
	\scriptsize
	
	\begin{tabular}{|c|c|c|c|c|c|c|c|c|c|c|c|} 
	\hline
	\multirow{2}{*}{Mod} &\multirow{2}{*}{Organ} & \multicolumn{5}{c|}{CNN} & \multicolumn{5}{c|}{SWIN} \\ 
	\cline{3-12} 
	{} &{} & {Rand} &{MG\cite{zhou2021models}} & {CPC\cite{taleb20203d}} &{Cub++\cite{tao2020revisiting}} & {PRCL\cite{zhou2021preservational}} & {Rand} & {DINO\cite{caron2021emerging}} &{iBOT\cite{zhou2022image}} & {SSIM\cite{xie2021simmim}} & {SMIT} \\ \hline
	\multirow{14}{*}{CT} &{Sp} &{0.930} &{0.950}&{0.940} &{0.926} &{0.937} &{0.944} &{0.946} & {0.948} & {0.950} & {\textbf{0.963}} \\
	{} &{RK} & {0.892} & {0.934} & {0.916} & {0.928} & {0.919} & {0.926}&{0.931}&{0.936}&{0.934}&{\textbf{0.950}} \\ 
	{} &{LK} & {0.894} & {0.918} & {0.903} & {0.914} & {0.921} & {0.905}&{0.913}&{0.919}&{0.913}&{\textbf{0.943}} \\
	{} &{GB} & {0.605} & {0.639} & {0.718} & {0.715} & {\textbf{0.797}} & {0.694}&{0.730}&{0.777}&{0.761}&{0.787} \\ 
	{} &{ESO} & {0.744} & {0.739} & {0.756} & {0.768} & {0.759} & {0.732}&{0.752}&{0.760}&{0.772}&{\textbf{0.772}} \\ 
	{} &{LV} & {0.947} & {0.967} & {0.953} & {0.946} & {0.954} & {0.950}&{0.954}&{0.956}&{0.956}&{\textbf{0.970}} \\ 
	{} &{STO} & {0.862} & {0.879} & {0.896} & {0.881} & {0.877} & {0.861}&{0.891}&{0.900}&{0.898}&{\textbf{0.903}} \\ 
	{} &{AOR} & {0.875} & {0.909} & {0.900} & {0.892} & {0.894} & {0.885}&{0.906}&{0.901}&{0.905}&{\textbf{0.913}} \\ 
	{} &{IVC} & {0.844} & {0.882} & {0.855} & {0.866} & {0.851} & {0.851}&{0.866}&{0.879}&{0.867}&{\textbf{0.871}} \\ 
	{} &{SPV} & {0.727} & {0.739} & {0.731} & {0.734} & {0.760} & {0.725}&{0.752}&{0.759}&{0.754}&{\textbf{0.784}} \\ 
	{} &{Pan} & {0.719} & {0.706} & {0.726} & {0.731} & {0.693} & {0.688}&{0.763}&{0.755}&{0.764}&{\textbf{0.810}} \\ 
	{} &{RA} & {0.644} & {0.671} & {0.655} & {0.665} & {0.661} & {0.660}&{0.651}&{0.659}&{0.640}&{\textbf{0.669}} \\ 
	{} &{LA} & {0.648} & {0.640} & {0.655} & {0.675} & {0.680} & {0.590}&{0.680}&{0.681}&{0.678}&{\textbf{0.687}} \\ 
    \cline{2-12} 
	{}&{\textcolor{black}{AVG.}}& {0.795} & {0.813} & {0.816} & {0.819} & {0.823} & {0.801}&{0.826}&{0.833}&{0.830}&{\textbf{0.848}} \\
	\hline
	
		\multirow{7}{*}{MR}  &{LV} & {0.921} &{0.936}&{0.925} &{0.920} &{0.930} &{0.922} &{0.920} & {0.939} & {0.937} & {\textbf{0.942} }\\
	{} &{LB} & {0.786} &{0.824}&{0.824} &{0.813} &{0.823} &{0.818} &{0.804} & {0.833} & {0.835} & {\textbf{0.855}} \\
	{} &{SB} & {0.688} &{0.741}&{0.745} &{0.735} &{0.745} &{0.708} &{0.729} & {0.744} & {0.759} & {\textbf{0.775}} \\
	{} &{STO} & {0.702} &{0.745}&{0.769} &{0.783} &{0.793} &{0.732} &{0.750} & {0.783} & {0.775} & {\textbf{0.812}} \\
	{} &{LK} & {0.827} &{0.832}&{0.876} &{0.866} &{0.876} &{0.837} &{0.911} & {0.883} & {0.874} & {\textbf{0.936}} \\
	{} &{RK} & {0.866} &{0.886}&{0.863} &{0.861} &{0.871} &{0.845} &{0.896} & {0.906} & {0.871} & {\textbf{0.930}} \\ 
    \cline{2-12} 
	{}&{\textcolor{black}{AVG.}}& {0.798} & {0.827} & {0.834} & {0.830} & {0.840} & {0.810}&{0.835}&{0.848}&{0.842}&{\textbf{0.875}} \\
	\hline

	\end{tabular}} 
\end{table*}

	\begin{figure*}[tb]
		\begin{center}
			\includegraphics[width=0.85\columnwidth,scale=1]{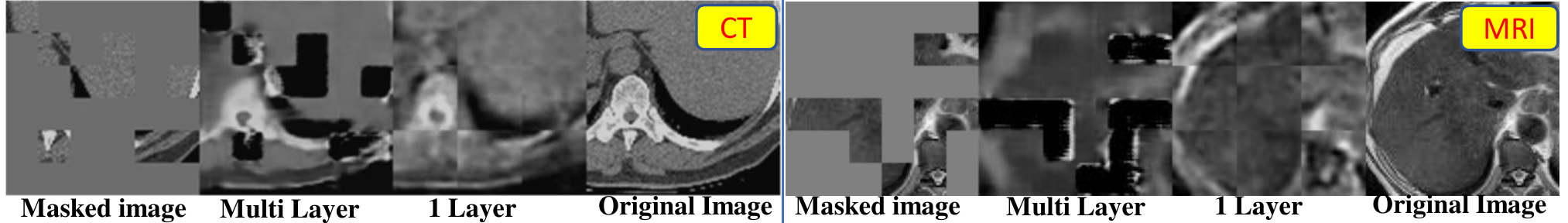} 
			\vspace{-0.05cm}\setlength{\belowcaptionskip}{-0.8cm}\setlength{\abovecaptionskip}{0.08cm}\caption{\small Reconstructed images using 1-layer vs. multi-layer decoder trained with SMIT from masked images (0.7 masking ratio).} \label{fig:rec_img}
		\end{center}
	\end{figure*}




\section{Discussion and conclusion}
In this work, we demonstrated the potential for SSL with 3D transformers for medical image segmentation. Our approach, which leverages CT volumes arising from highly disparate body locations and diseases showed feasibility to produce robustly accurate segmentations from CT and MRI scans and surpassed multiple current SSL-based methods, especially for hard to segment organs with high appearance variability and small sizes. Our introduced masked image dense prediction pretext task improved the ability of self distillation using MIM to segment a variety of organs from CT and MRI and with lower requirement of fine tuning dataset size. Our method shows feasibility for medical image segmentation. 
		
\section{Conclusion}
\bumpup

{\tiny
	\bibliographystyle{splncs}
}
\bibliography{mybibliography}
%
%
%
%





\end{document}